\documentstyle[12pt,psfig]{article}
\begin{document}
\begin{titlepage}
\begin{center}

{\Large\bf{Constraining the free parameter of the high parton density effects}}
\\[5.0ex]
{\Large\it{ M. B. Gay  Ducati $^{1 \,\,*}$\footnotetext{$^{*}$E-mail:gay@if.ufrgs.br}}}
 {\it and}
{ \Large \it{ V. P. Gon\c{c}alves $^{1,2 \,\,**}$\footnotetext{$^{**}$E-mail:barros@if.ufrgs.br; victor@inf.uri.br} 
}} \\[1.5ex]
{\it $^{1}$ Instituto de F\'{\i}sica, Univ. Federal do Rio Grande do Sul}\\
{\it Caixa Postal 15051, 91501-970, Porto Alegre, RS, BRAZIL}\\
\vspace{1cm}
{\it ${^{2}}$ Departamento de Ci\^encias Exatas e da Terra, Univ. Regional Integrada do Alto Uruguai e das Miss\~oes}\\
{\it CEP 98400-000, Frederico Westphalen, RS, BRAZIL}\\[5.0ex]
\end{center}

{\large \bf Abstract:}
The high parton  density  effects are strongly dependent of the spatial gluon
distribution within the proton, with  radius $R$, which cannot be derived from
perturbative QCD. In this paper we assume that the unitarity corrections are
present in the HERA kinematical region and constrain the value of  $R$ using
the data for the proton structure function and its slope. We obtain  that the
gluons are not distributed uniformly in the whole proton disc, but behave as
concentrated in smaller regions.

 \vspace{1.5cm}
 


\end{titlepage}

The description of the dynamics at high density parton regime is one of the  main open questions 
of the strong interactions theory. While in the region of moderate Bjorken $x$ ($x \ge 10^{-2}$) 
the  well-established methods of 
operator product expansion and renormalization group equations  have been applied successfully, the small $x$ region still lacks a consistent theoretical framework (For a review see \cite{cooper}). Basically, it is questionable the use of   the DGLAP equations \cite{dglap}, which reflects the dynamics at moderate $x$, in the region of small values of $x$, where the gluon distribution determines the behavior of the observables. The traditional procedure of using the DGLAP equations  to calculate the gluon distribution at small $x$ and large momentum transfer $Q^2$ is by summing the leading powers of  $\alpha_s\,ln\,Q^2\,ln(\frac{1}{x})$, where $\alpha_s$ is  the strong coupling constant, known as   the double-leading-logarithm approximation (DLLA). In axial gauges, these leading double logarithms are generated by ladder diagrams in which the emitted gluons have strongly ordered transverse momenta, as well as strongly ordered longitudinal momenta. Therefore the DGLAP must breakdown at small values of $x$, firstly because this framework does not account for the  contributions to the cross section which are leading  in $\alpha_s \, ln (\frac{1}{x})$ \cite{bfkl}. Secondly, because the parton densities  become large and there is  need to develop a high density formulation of QCD \cite{hdqcd}, where the unitarity corrections are considered.

There has been intense debate on to which extent non-conventional QCD evolution is required by the deep inelastic $ep$ HERA data \cite{cooper}. Good fits to the $F_2$ data for $Q^2 \ge 1\,GeV^2$ can be obtained from distinct approaches, which consider DGLAP and/or BFKL evolution equations \cite{ball,martin}.  In particular, the conventional  perturbative QCD approach is very successful in describing the main features of HERA data and, hence, the signal of a new QCD dynamics has been in general hidden or mimicked by a strong background of conventional QCD evolution. For instance, recently the magnitude of the higher twist terms was demonstrated to be large in the transverse and longitudinal structure function, but as these  contributions have opposite signal the effect in the behavior of $F_2$ structure function is small \cite{barbon}. At this moment there are some possible signs  of the high density dynamics in the HERA kinematical region: the behavior of the slope of  structure function and the energy dependence of diffractive structure functions \cite{muelec}. However,  more studies and precise data are still needed.

Our goal in this letter is by assuming the presence of the high density effects in the HERA data, to  constrain the spatial distribution of the gluons inside the proton. The radius $R$ is a phenomenological parameter which is not present in the linear dynamics (DGLAP/BFKL) and is introduced when the unitarity corrections are estimated. In general, the evolution equations at high density QCD (hdQCD) \cite{hdqcd} resum  the powers of the function 
$\kappa (x,Q^2) \equiv \frac{\alpha_s N_c  \pi }{2 Q^2 R^2} xG(x,Q^2)$, 
which represents the probability  of gluon-gluon interaction inside the parton cascade. At this moment, these evolution equations 
\begin{itemize}
\item match the DLA limit of the DGLAP  evolution equation in the limit of low parton densities $(\kappa \rightarrow 0)$;
\item match the GLR equation  in first order of unitarity corretions [${\cal{O}}(\kappa^2)$]. 
\end{itemize}
Although the complete demonstration of the  equivalence between these formulations in the region of large $\kappa$ is  still an open question, some steps in this direction were given recently \cite{npbvic1,npbvic2}.  One of the main characteristics in common of these approaches is the behavior of the   structure function in the asymptotic regime of very high density \cite{npbvic2}: $F_2 (x, Q^2) \propto Q^2 R^2 ln (1/x)$. Therefore, although the parameter $R$ cannot be derived from perturbative QCD,   the unitarity corrections  crucially depend on its value. 
Here we will  discriminate the range of possible values of $R$ considering  the AGL approach for the high density systems and  the HERA data for the structure function $F_2(x,Q^2)$ and its slope. In the HERA kinematical region the solution of the AGL equation is identical to the GLR solution \cite{ayala2}, which implies  that our estimates in principle are not model dependent.

We will consider initially the physical interpretation of the $R$ parameter, and   present later   a brief review of the AGL approach and our estimates. The value of $R$ is associated with the coupling of  the gluon ladders with the proton, or to put it in  another way, on how the gluons are distributed within the proton. $R$ may be of  the order of the proton radius if the gluons are distributed uniformly in the whole proton disc or much smaller
if the gluons are concentrated, {\it i.e.} if the gluons in the proton are confined in a disc with smaller radius than the size of the proton \cite{hotspot}. In a first approximation,  the radius is  expected to be smaller than the proton radius. This affirmative is easy to understand.
Consider the first order contribution to the unitarity corrections presented in  Fig. \ref{fig1}, where two  ladders couple to the proton.  The  ladders may be attached to  different constituents of the proton or to the same constituent. In the first case [Fig. \ref{fig1} (a)] the unitarity corrections are controlled by the proton radius, while in the second case [Fig. \ref{fig1} (b)] these corrections are controlled by the constituent radius, which is  smaller than the proton radius.   Therefore, on the average, we expect that the radius will be smaller than the proton radius. The value of  $R^2$ 
reflects the integration over $b_t$ in the first diagrams for the unitarity corrections.
 
In our estimates for the parameter $R$ we will use the AGL approach \cite{ayala2}. Here we present only a brief review of this approach and refer the original papers for details. Basically, in the AGL approach 
the behavior of the gluon distribution can be obtained from the cross section for the interaction of a virtual gluon with a proton. In the target rest frame the  virtual gluon at high energy
(small $x$) decay into a gluon-gluon pair long before the 
interaction with the target. The $gg$ pair subsequently interacts 
with the target, with  the transverse distance $r_t$ between the gluons  assumed fixed. It allows to factorize the total 
cross section between the wave function of the virtual gluon and the interaction 
cross section of the gluon-gluon  pair with the target. The gluon wave function 
is calculable and the interaction cross section is modelled. Considering the unitarity  corrections for the  interaction cross section results that the gluon distribution is given by the Glauber-Mueller formula \cite{ayala2}  
\begin{eqnarray}
xG(x,Q^2) = \frac{2R^2}{\pi^2}\int_x^1 \frac{dx^{\prime}}{x^{\prime}}
\int_{\frac{1}{Q^2}}^{\frac{1}{Q_0^2}} \frac{d^2r_t}{\pi r_t^4} \{C 
+ ln(\kappa_G(x^{\prime}, \frac{1}{ r_t^2})) + E_1(\kappa_G(x^{\prime}, \frac{1}{r_t^2}))\}  \,\,,
\label{masterg}
\end{eqnarray} 
where $C$ is the Euler constant,  $E_1$ is the exponential function and  the function  $\kappa_G(x, \frac{1}{r_t^2}) = \frac{3 \alpha_s}{2R^2}\,\pi\,r_t^2\, xG(x,\frac{1}{r_t^2})$. If equation (\ref{masterg}) is expanded for small $\kappa_G$, 
 the first term (Born term) will correspond to 
the usual DGLAP equation in the small $x$ region, while 
 the other terms will take into account the unitarity corrections.
The Glauber-Mueller formula is a particular case of the AGL equation proposed in Ref. \cite{ayala2}, and a good approximation for the solutions of this equation in  the HERA kinematical region, which we will use in this work.

A similar procedure can be used to estimate the unitarity corrections for the $F_2$ structure function and its slope. In the target rest frame the 
proton structure function is given by \cite{nik}
\begin{eqnarray}
F_2(x,Q^2) = \frac{Q^2}{4 \pi \alpha_{em}} \int dz \int d^2r_t |\Psi(z,r_t)|^2 \, \sigma^{q\overline{q}}(z,r_t)\,\,,
\label{f2target}
\end{eqnarray}
where $\Psi$ is the photon wave function and the cross section $\sigma^{q\overline{q}}(z,r_t^2)$ describes the interaction of the pair with the target. Considering only light quarks ($i=u,\,d,\,s$), the expression for $\Psi$ derived in Ref. \cite{nik} and the unitarity corrections for the interaction cross section of the $q\overline{q}$ with the proton,  the    $F_2$ structure function can be written in the AGL approach as \cite{ayala2}
\begin{eqnarray}
F_2(x,Q^2) =  \frac{R^2}{2\pi^2} \sum_{f=u,d,s} e_f^2 \int_{\frac{1}{Q^2}}^{\frac{1}{Q_0^2}} \frac{d^2r_t}{\pi r_t^4} \{C + ln(\kappa_q(x, r_t^2)) + E_1(\kappa_q(x, r_t^2))\}\,\,,
\label{diseik2}
\end{eqnarray}
where the function  
$\kappa_q(x, r_t^2) = 4/9 \kappa_G (x,r_t^2)$. 
The  slope of $F_2$ structure function in this approach is straightforward from the expression (\ref{diseik2}). We obtain that
\begin{eqnarray}
\frac{dF_2(x,Q^2)}{dlogQ^2} =  \frac{R^2 Q^2}{2\pi^2} \sum_{u,d,s} e_f^2 
 \{C + ln(\kappa_q(x, Q^2)) + E_1(\kappa_q(x, Q^2))\}\,\,.
\label{df2eik}
\end{eqnarray}

The expressions (\ref{diseik2}) and (\ref{df2eik}) predict  the behavior of the unitarity corrections to $F_2$ and its slope considering the AGL approach for the interaction of the $q\overline{q}$ with the target.  In this case we are calculating  the corrections associated 
with the crossing of the $q\overline{q}$ pair through the target. Following \cite{glmn} 
we will denote this contribution as the quark sector contribution to the unitarity corrections. 
However, the behavior of $F_2$ and its slope are associated with the behavior of the gluon distribution used as input in  (\ref{diseik2}) and (\ref{df2eik}). In general, it is assumed that the gluon distribution is described by a parametrization of the parton distributions (for example: GRV, MRS, CTEQ). In this case the unitarity corrections 
in the gluon distribution are not included explicitly. However,  calculating the unitarity corrections using the AGL approach we obtain that they imply large corrections in the behavior of the gluon distribution  in the HERA kinematical region, and therefore cannot be disregarded. Therefore, we should consider the solution from Eq. (\ref{masterg}) as input in the Eqs. (\ref{diseik2}) and (\ref{df2eik}) in order to accurately determine   the behavior of $F_2$ and its slope, {\it i.e.}, we should consider the unitarity corrections  in the quark and the gluon sectors (quark+ gluon sector).

The expressions  (\ref{masterg}), (\ref{diseik2}) and  (\ref{df2eik})  are correct in the double leading logarithmic approximation (DLLA). As shown in \cite{ayala2} the DLLA does not work quite well in the whole  accessible kinematic region ($Q^2 > 
0.4 \,GeV^2$ and $x > 10^{-6}$). Consequently, a more realistic approach must 
be considered to calculate the observables in the HERA kinematical region. In \cite{ayala2} the subtraction of the Born term and the addition of the GRV parametrization \cite{grv95} were proposed to the $F_2$ case. In this case we have
\begin{eqnarray}
F_2(x,Q^2) =  F_2(x,Q^2) \mbox{[Eq.  (\ref{diseik2})]} -  F_2(x,Q^2) 
\mbox{[Born]} +  F_2(x,Q^2) \mbox{[GRV]} \,\,\, ,
\label{f2}
\end{eqnarray}
where the Born term is the first term  in the expansion in $\kappa_q$ 
of the equations  (\ref{diseik2})(see \cite{prd} 
for more details). Here we apply this procedure for the $F_2$ slope. In this case 
\begin{eqnarray}
\frac{dF_2(x,Q^2)}{dlogQ^2} =  \frac{dF_2(x,Q^2)}{dlogQ^2} \mbox{[Eq.  
(\ref{df2eik})]} -  \frac{dF_2(x,Q^2)}{dlogQ^2} \mbox{[Born]} +  
\frac{dF_2(x,Q^2)}{dlogQ^2} \mbox{[GRV]} \,\,\, ,
\label{df2}
\end{eqnarray}
where the Born term is the first term in the expansion in $\kappa_q$  of the 
equation (\ref{df2eik}). The last term is associated with  
the traditional DGLAP framework, which 
at small values of $x$ predicts 
\begin{eqnarray}
\frac{dF_2(x,Q^2)}{dlogQ^2} = \frac{10 \alpha_s(Q^2)}{9 \pi} \int_0^{1-x} 
dz \, P_{qg}(z) \, \frac{x}{1-z}g\left(\frac{x}{1-z},Q^2\right)\,\,,
\label{df2glap}
\end{eqnarray}
where $\alpha_s(Q^2)$ is the  running  coupling  constant  and the splitting function $P_{qg}(x)$ gives the probability to find a quark with momentum fraction $x$ inside a gluon. This equation describes the scaling violations of the proton structure function in terms of the gluon distribution. We use the GRV parametrization as input in the expression (\ref{df2glap}).


 In Fig. \ref{fig2} we compare our predictions for unitarity corrections in the $F_2$ structure 
function and the H1 data \cite{h1} as a function of $ln(\frac{1}{x})$ at different virtualities and some values of the radius.  We see clearly that the unitarity  corrections strongly increase at small values of the radius $R$. Our goal is not a best fit, but to eliminate some values of radius  comparing the predictions of the AGL approach and  HERA data.  The choice $R^2 = 1.5 \, GeV^2$  does not describe the data, {\it i.e.} the data discard the possibility of  very large SC in the HERA kinematic region. However, there are still a large range  ($ 5 \le R^2 \le 12$) of possible values  for the radius which reasonably describe the  $F_2$ data.  To discriminate 
between these possibilities we must consider the behavior of the  $F_2$ slope. 

In Fig. \ref{fig3} (a) we present our results for  $\frac{dF_2(x,Q^2)}{dlogQ^2}$ considering initially the unitarity corrections only in the quark sector, { \it i.e.} using the GRV parameterization for the gluon distribution as input in Eq. (\ref{df2eik}). The ZEUS data  points \cite{zeus} correspond to  different $x$ and $Q^2$ value. The 
$(x,Q^2)$ points are averaged values obtained from each of the experimental 
data distribution bins.   Only the data points with $<Q^2> \, \ge 0.52\,GeV^2$ 
and $x < 10^{-1}$ were used here. 
 Our results show that the fit to the data occurs at small values of $R^2$,  which  are discarded by the $F_2$ data. Therefore, in agreement with our previous conclusions, we must consider the unitarity corrections to the gluon distribution to describe consistently the $F_2$ and $\frac{dF_2(x,Q^2)}{dlogQ^2}$ data. In Fig. \ref{fig3} (b) we present our results for  $\frac{dF_2(x,Q^2)}{dlogQ^2}$ considering the SC in the gluon and quark sectors for different values of $R^2$, calculated using the AGL approach. The best result occurs for $R^2 = 5\,GeV^{-2}$, which also describes the $F_2$ data.

 The value for the squared radius $R^2 = 5\,GeV^{-2}$ obtained in our analysis agrees with the estimates obtained using the  HERA data on diffractive photoproduction of $J/\Psi$ meson \cite{zeusjpsi,h1jpsi}. Indeed, the experimental values for the slope are $B_{el} = 4 \, GeV^{-2}$ and $B_{in} = 1.66\,GeV^{-2}$ and the cross section for $J/\Psi$ diffractive production with and without photon dissociation are equal. Neglecting the $t$ dependence of the pomeron-vector meson coupling  the value of $R^2$ can be estimated \cite{plb}. It turns out that $R^2 \approx 5\,GeV^{-2}$, {\it i.e.}, approximately 2 times smaller than the radius of the proton.

Some additional comments are in order. The unitarity corrections to $F_2$ and 
its slope may also be analysed using a two radii model for the proton \cite{glmn1}. Such analysis is motivated by the large difference between the measured slopes in elastic and inelastic diffractive leptoproduction of vector mesons in DIS. An analysis using the   two radii model for the proton is not a goal  of this paper, since a final conclusion on the correct model is still under debate. 

The AGL approach describe the ZEUS data,  
as well as the DGLAP evolution equations using  modified parton distributions. 
Recently, the MRST  \cite{mrst} and GRV group \cite{grv98} have proposed a different set of parton 
parametrizations which considers an initial 'valence-like' gluon distribution. 
In Fig. \ref{fig4} we present the predictions of the DGLAP  dynamics using the GRV(98) parameterization as input. We can see that this
  parametrization allows to describe the $F_2$ slope data without an unconventional effect. This occurs because there is  a large freedom in the initial parton distributions and the initial virtuality used in these parametrization, demonstrated by the large difference between the predictions obtained using the GRV(94) or GRV(98) parametrization. In our analysis we assume that the unitarity corrections are present in the HERA kinematical region, mainly in the $F_2$ slope, which is directly dependent of the behavior of the gluon distribution. Therefore, we consider that the fail of the DGLAP evolution equation plus GRV(94) to describe the ZEUS data as an evidence of the high density effects and the possibility of description of the data using new parametrizations as a way to hidden these effects. In Fig. \ref{fig4} we also show that  if the GRV(98) parameterization is used as input in the calculations of the high density effects in the HERA kinematical region, the ZEUS data cannot be described.

A comment related to the $F_2$ slope HERA data is important. The ZEUS data \cite{zeus} were obtained in a limited region of the phase space. Basically, in these data a value for the $F_2$ slope is given for a pair of values of $x$ and $Q^2$, {\i.e.} if we plot the $F_2$ slope for fixed $Q^2$ as a function of $x$ we have only one data point in the graphic. Recently,  the H1 collaboration has presented a preliminary set of data for the $F_2$ slope \cite{klein} obtained in a large region of the phase space. The main point is that these new data allows us an analysis of the behavior of the $F_2$ slope as a function of $x$ at fixed $Q^2$. In Fig. \ref{fig5} we present the comparison between the predictions of the DGLAP approach, using the GRV(94) or GRV(98) parametrization as input, and the AGL approach with the preliminary H1 data [extracted from the Fig. 13 of \cite{klein}]. We can see that, similarly to observed in the ZEUS data, the DGLAP + GRV(94) prediction cannot describe the data, while the AGL approach describe very well this set of data. We believe that our conclusions are not modified if these new data in a large phase space are included in the analysis made in this letter.

  In this paper we have assumed that the unitarity corrections (high density effects) are present in the HERA kinematical region and believe that only   a comprehensive  analysis of distinct observables ($F_L, \, F_2^c, \, \frac{dF_2(x,Q^2)}{dlogQ^2}$)  will allow a more careful evaluation of the unitarity corrections at small $x$ \cite{prd,ayala3}.
 The  main conclusion of this paper is that the analysis of the $F_2$ and $\frac{dF_2(x,Q^2)}{dlogQ^2}$ data using the AGL approach implies that the gluons are not distributed uniformly in the whole proton disc, but behave as concentrated in smaller regions. This conclusion motivates an analysis of the jet production, which probes smaller regions within the proton, in a calculation  that includes the high density effects.

\section*{Acknowledgments}
This work was partially financed by Programa de Apoio a N\'ucleos de Excel\^encia (PRONEX) and CNPq, BRAZIL.

\newpage
\section*{Figure Captions}

\begin{figure}
\centerline{\psfig{file=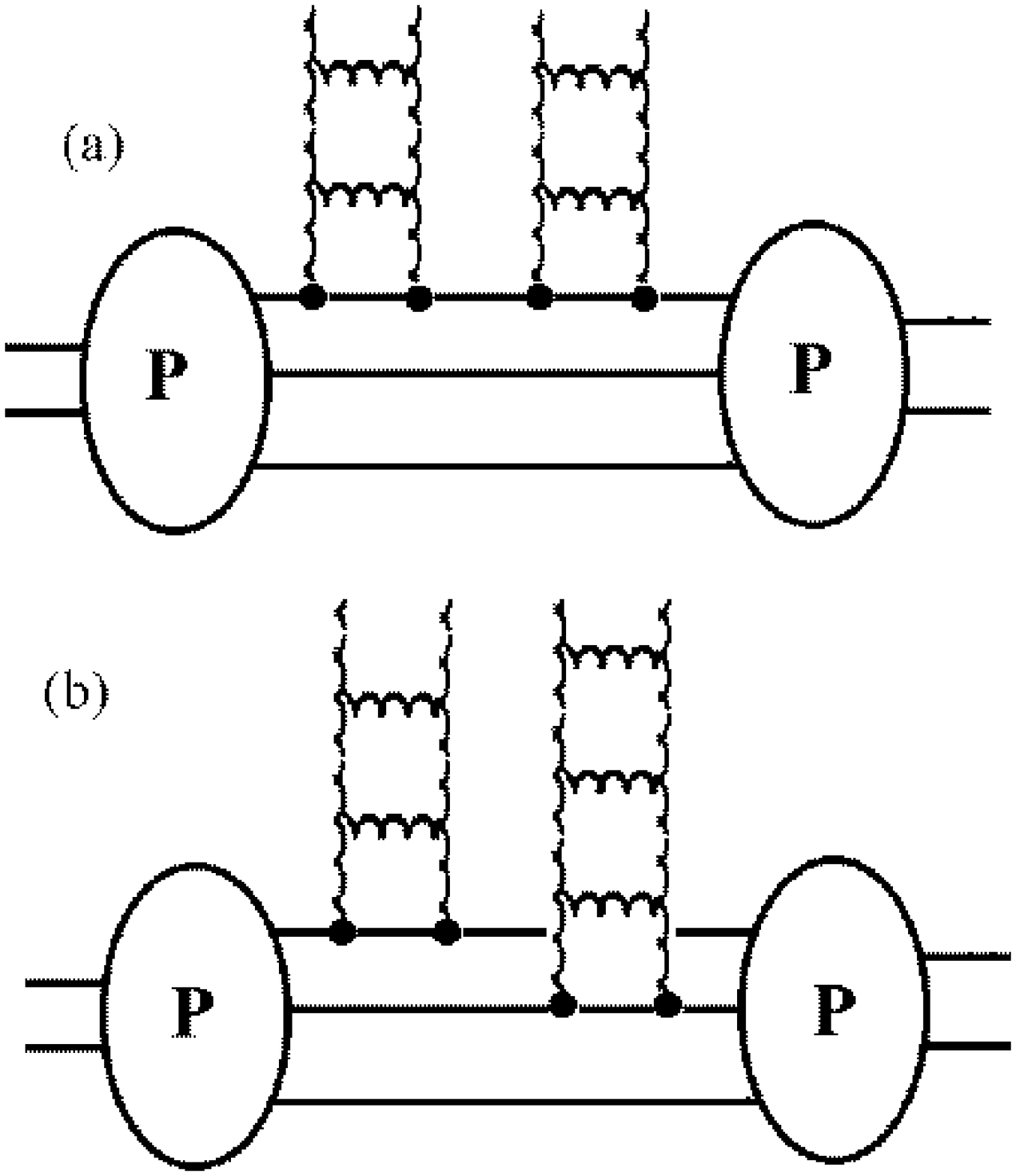,width=100mm}}
\caption{}
\label{fig1}
\end{figure}

\vspace{1.0cm}
Fig. \ref{fig1}: First order contribution to the unitarity corrections. In  (a) these corrections are controlled by the proton radius, while in (b) by the constituent radius.


\vspace{1.0cm}

\begin{figure}
\centerline{\psfig{file=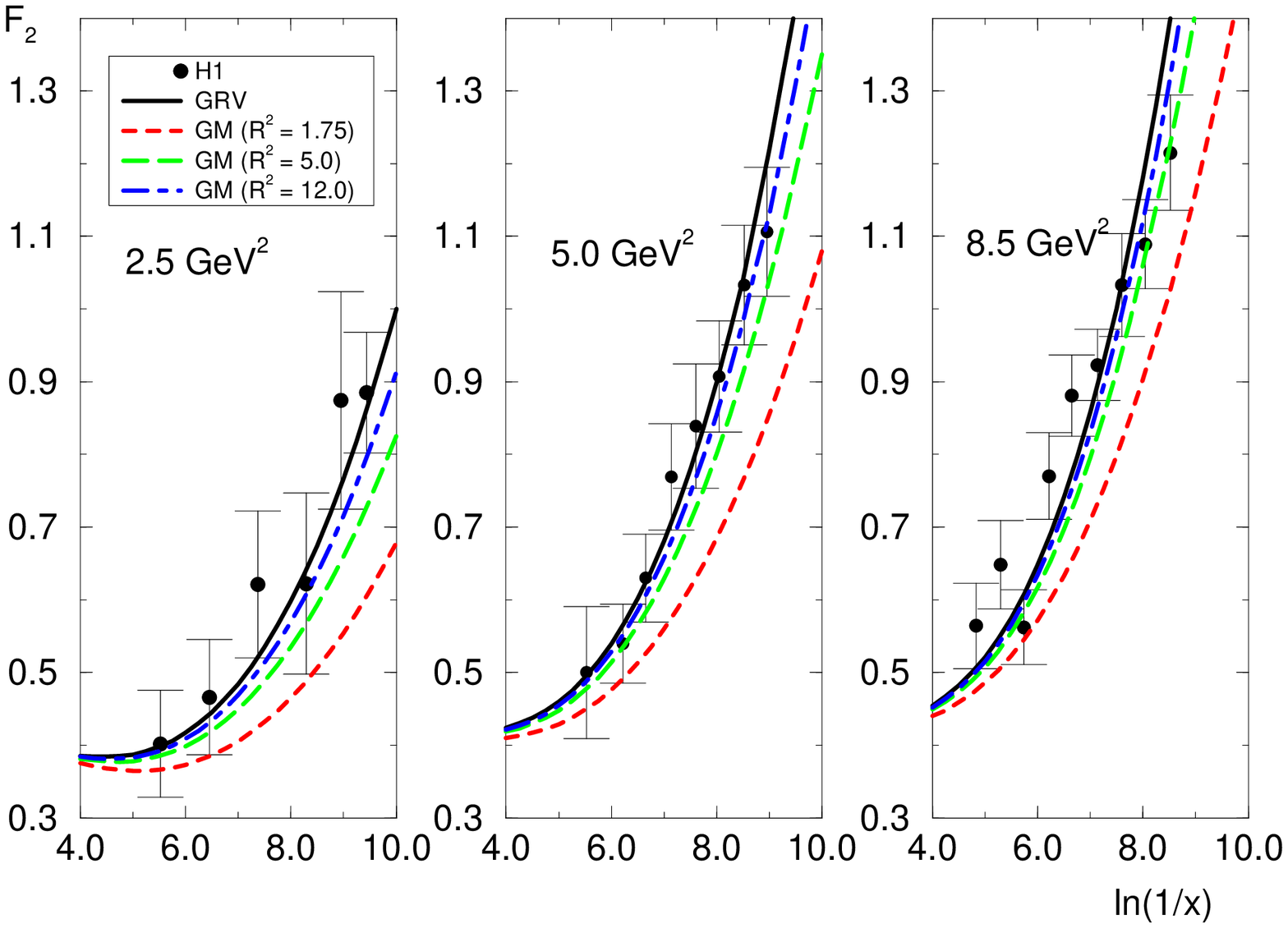,width=150mm}}
\caption{ }
\label{fig2}
\end{figure}

Fig. \ref{fig2}: The $F_2$ structure function as a function of the variable $ln(\frac{1}{x})$ 
for different virtualities and radii.  Only the unitarity  corrections in the quark sector 
are considered. Data from H1 \cite{h1}.

\vspace{1.0cm}

\begin{figure}[t]
\begin{tabular}{c c}
\psfig{file=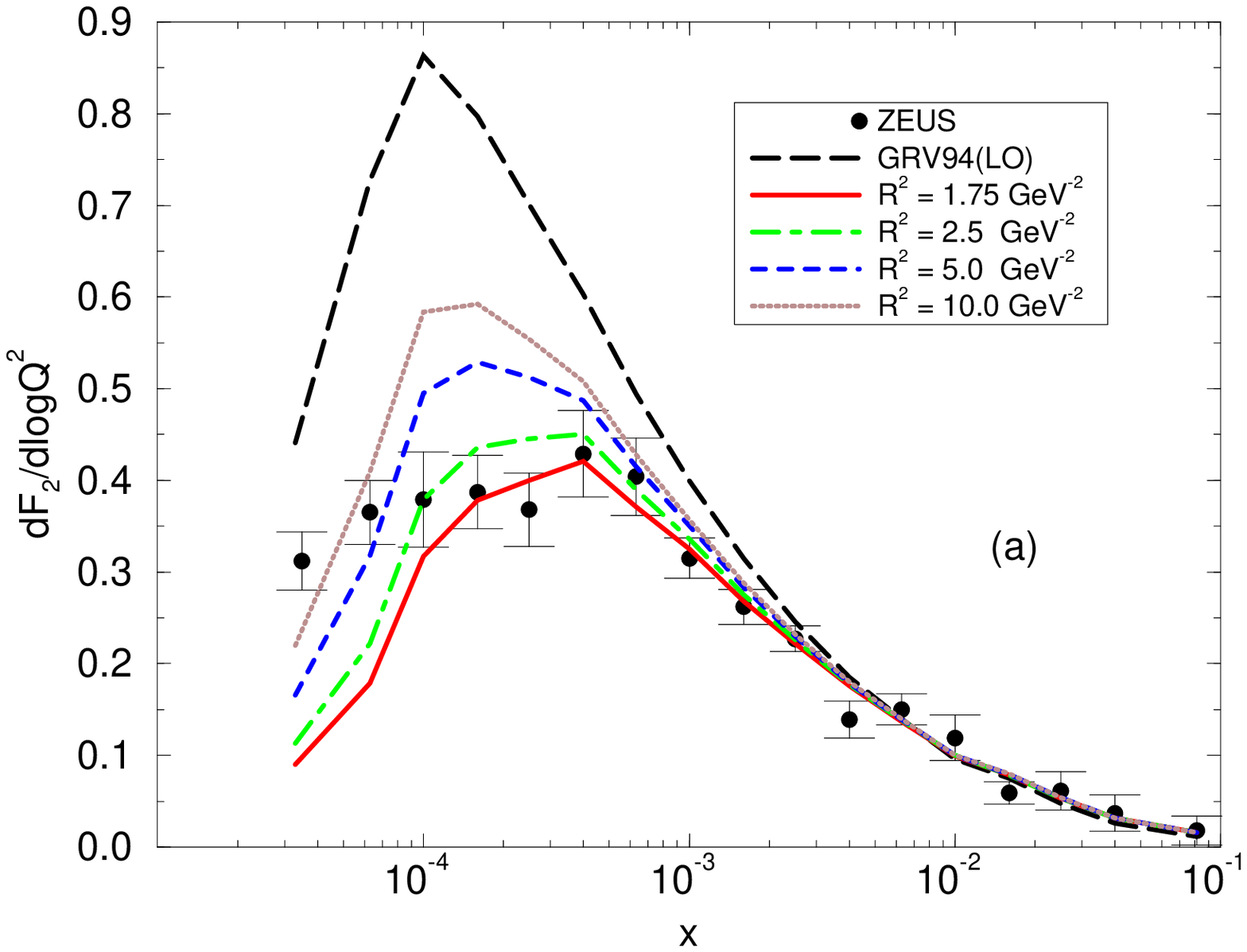,width=100mm} \\ \psfig{file=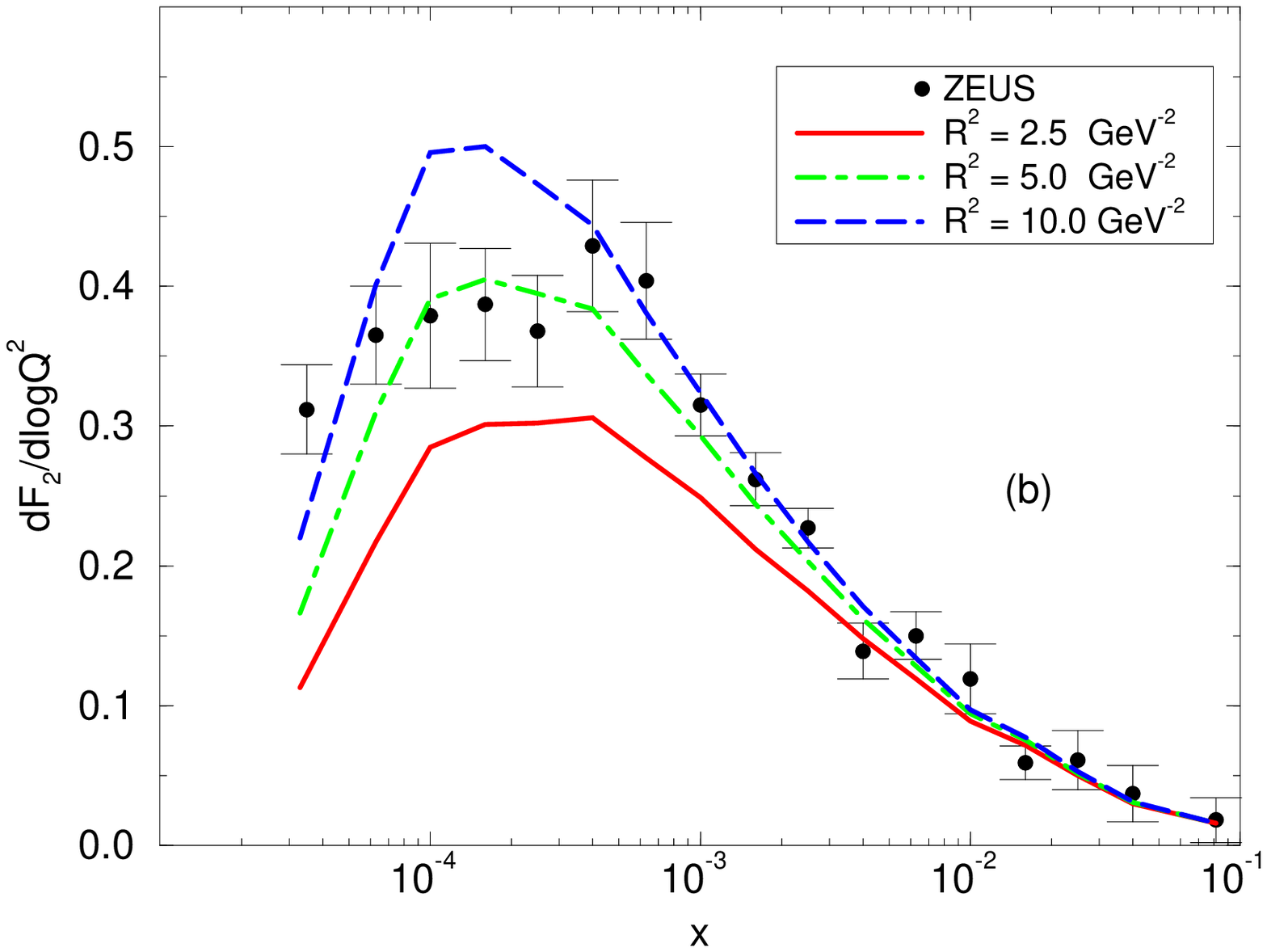,width=100mm} \end{tabular}
\caption{}
\label{fig3}
\end{figure}

Fig. \ref{fig3}: The $F_2$  slope  as a function of the variable $x$ for different radii. 
(a) Only the unitarity corrections in the quark sector are considered.  
  (b) The unitarity corrections in the gluon-quark sector are considered.  Data from 
ZEUS \cite{zeus}. The data points correspond to a different $x$ and $Q^2$ value.

\vspace{1.0cm}

\begin{figure}
\centerline{\psfig{file=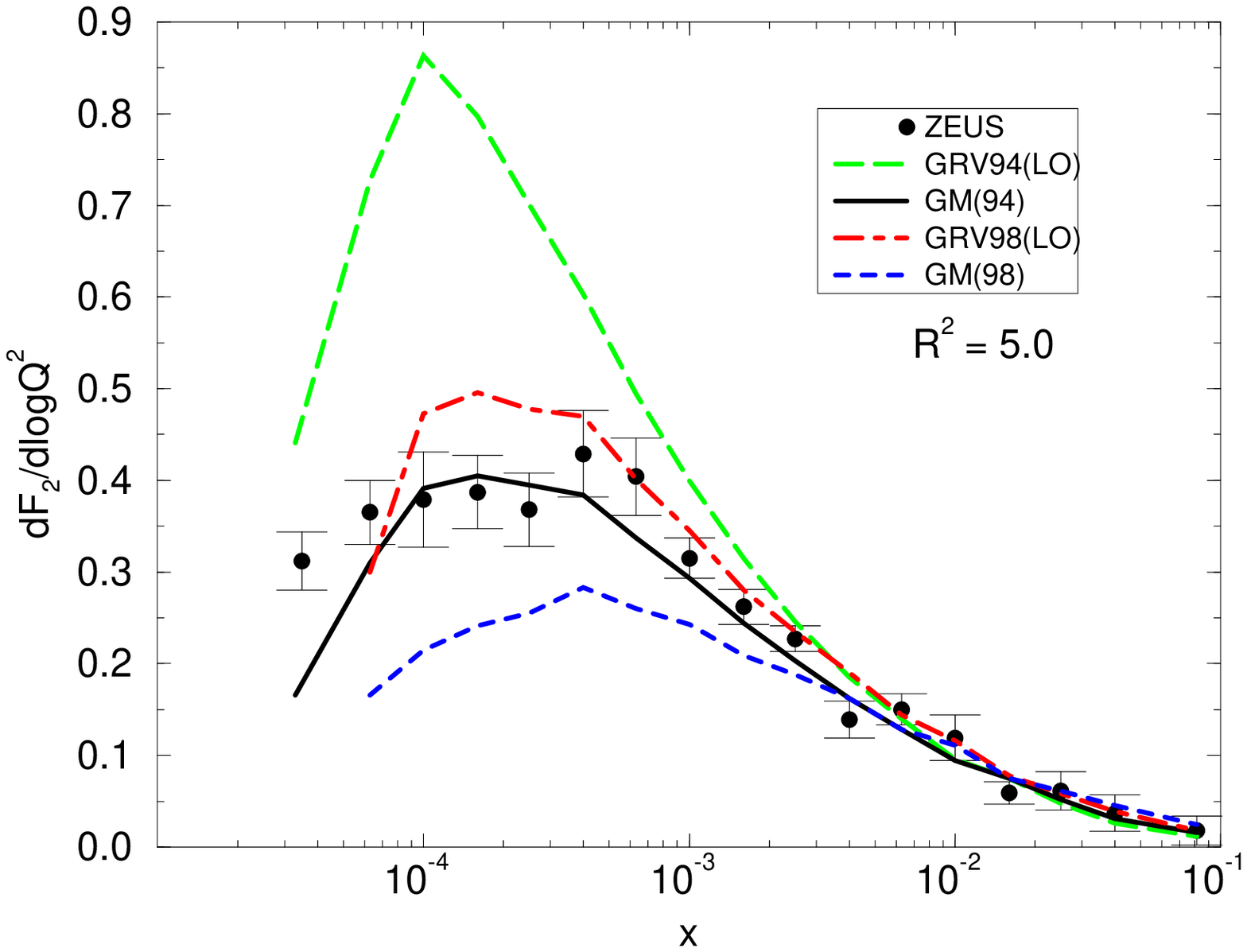,width=150mm}}
\caption{ }
\label{fig4}
\end{figure}

Fig. \ref{fig4}: Comparison between the DGLAP and Glauber-Mueller (GM) predictions for the behavior of the $F_2$ slope  using as input in the calculations the GRV(94) or GRV(98) parameterizations.  Data from 
ZEUS \cite{zeus}. The data points correspond to a different $x$ and $Q^2$ value.

\vspace{1.0cm}

\begin{figure}
\centerline{\psfig{file=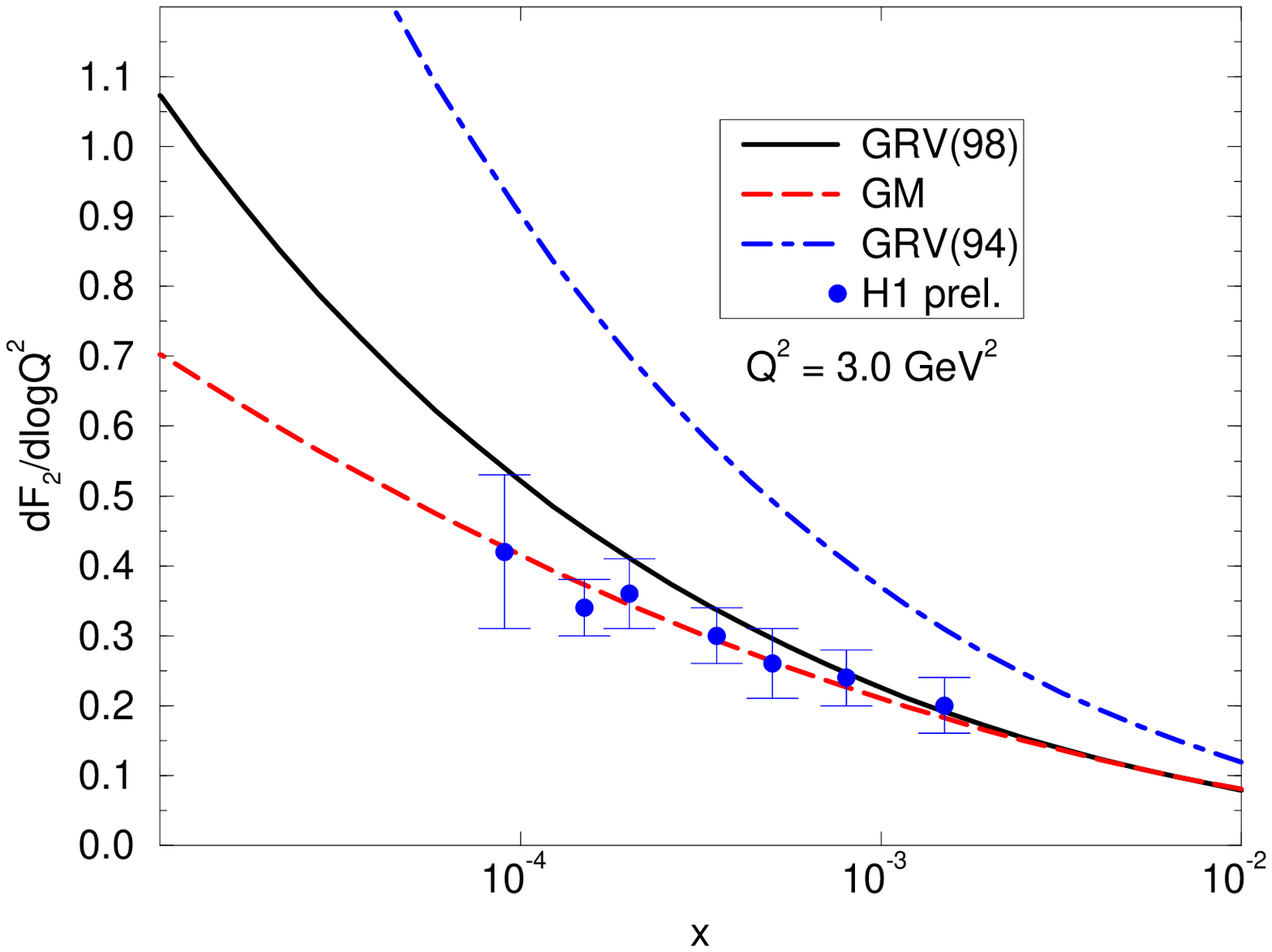,width=150mm}}
\caption{ }
\label{fig5}
\end{figure}

Fig. \ref{fig5}: Comparison between the Glauber-Mueller (GM) prediction and DGLAP  using as input in the calculations the GRV(94) or GRV(98) parameterizations, for the behavior of the $F_2$ slope.  Preliminary data from 
H1 \cite{klein}. 

\newpage

\begin{thebibliography}{99}

\bibitem{cooper}
A. M. Cooper-Sarkar, R.C.E. Devenish and A. De Roeck, 
{\sl Int. J. Mod. Phys} {\bf A13} (1998) 3385.



\bibitem{dglap}

 Yu. L. Dokshitzer. {\sl Sov. Phys. JETP} {\bf 46} (1977) 641;
 G. Altarelli and G. Parisi. {\sl Nucl. Phys.} {\bf B126} (1977) 298;
 V. N. Gribov and L.N. Lipatov. {\sl Sov. J. Nucl. Phys} {\bf 15} (1972) 438.



\bibitem{bfkl}
 E.A. Kuraev, L.N. Lipatov and V.S. Fadin. {\sl Phys. Lett} {\bf B60} (1975) 50; {\sl Sov. Phys. JETP} {\bf 44} (1976) 443; {\sl Sov. Phys. JETP} {\bf 45} (1977) 199;
    Ya. Balitsky and L.N. Lipatov. {\sl Sov. J. Nucl. Phys. } {\bf 28} (1978) 822.




\bibitem{hdqcd}
 L. V. Gribov,
  E. M. Levin, M. G. Ryskin. {\sl Phys.Rep.} {\bf 100} (1983) 1; A. L. Ayala, M. B. Gay Ducati, E. M. Levin.  {\sl Nucl. Phys.} {\bf B493} (1997) 305; J. Jalilian-Marian  {\it et al.} {\sl Phys. Rev.} {\bf D59} (1999) 034007;
Y. U. Kovchegov.{\sl Phys Rev.} {\bf D60} (1999) 034008.



\bibitem{ball}
 R. Ball and S. Forte {\sl Phys. Lett} {\bf B335} (1994) 77.






\bibitem{martin} 
J. Kwiecinski, A. D. Martin and A. M. Stasto,  {\sl  Phys. Rev.} {\bf D56} (1997) 3991.


\bibitem{barbon}
J. Bartels, C. Bontus,  {\sl  Phys. Rev.} {\bf D61} (2000) 034009.


\bibitem{muelec}
A. H. Mueller. hep-ph/9911289.


\bibitem{npbvic1}
M. B. Gay Ducati, V. P. Gon\c{c}alves. {\sl Nuc. Phys.} {\bf B557} (1999) 296.


\bibitem{npbvic2}
M. B. Gay Ducati, V. P. Gon\c{c}alves.  hep-ph/0003098.


\bibitem{ayala2}
A. L. Ayala, M. B. Gay Ducati and E. M. Levin,  
 {\sl Nucl. Phys.} {\bf B511} (1998) 355.



\bibitem{hotspot} 
A. H. Mueller,  {\sl Nucl. Phys. B (Proc. Suppl.)} {\bf 18C} (1990) 125;
J. Bartels, A. De Roeck and M. Loewe,  {\sl Z. Phys.} {\bf C54} (1992) 635.


\bibitem{mueller}
A. H. Mueller,  {\sl Nucl. Phys.} {\bf B335} (1990) 115.



\bibitem{nik} 
N. Nikolaev and B. G. Zakharov,  {\sl Z. Phys.} {\bf C49} (1990) 607.


\bibitem{plb}
A. L. Ayala, M. B. Gay Ducati and E. M. Levin, {\sl Phys. Lett.} {\bf B388} (1996) 188.


\bibitem{glmn} 
E. Gotsman {\it et al.},  {\sl Nucl. Phys.} {\bf B539} (1999) 535.



\bibitem{grv95} 
M. Gluck, E. Reya and A. Vogt.  {\sl Z. Phys.} {\bf C67} (1995) 433.


\bibitem{prd}
A. L. Ayala, M. B. Gay Ducati and V. P. Gon\c{c}alves,  {\sl  Phys. Rev.} {\bf D59} (1999) 054010.


\bibitem{h1}
 S. Aid {\sl et al.}. {\sl Nucl. Phys.} {\bf B470} (1996) 3.


\bibitem{zeus} 
 M. Derrick {\sl et al.} (ZEUS COLLABORATION), {\sl Eur. Phys. J.} {\bf C7} (1999) 609.



\bibitem{zeusjpsi} 
 M. Derrick {\sl et al.} (ZEUS COLLABORATION),  {\sl Phys. Lett} {\bf B350} (1995) 120.


\bibitem{h1jpsi}
 S. Aid {\sl et al.} (H1 COLLABORATION), {\sl Nucl. Phys.} {\bf B472} (1996) 3.



\bibitem{glmn1} 
E. Gotsman, E. Levin and U. Maor.  {\sl Phys. Lett} {\bf B425}, 369 (1998).



\bibitem{mrst} 
A. D. Martin {\it et al.},  {\sl Eur. Phys. J.} {\bf C4} (1998) 463.




\bibitem{grv98} 
M. Gluck, E. Reya and A. Vogt,  {\sl Eur. Phys. J.} {\bf C5} (1998) 461.


\bibitem{klein}
M. Klein, hep-ex/0001059.

\bibitem{ayala3}
A. L. Ayala, M. B. Gay Ducati and E. M. Levin,  {\sl Eur. Phys. J.} {\bf C8} (1999) 115.



\end{thebibliography}
\end{document}